# Taxonomy and synthesis of Web services querying languages

Ibrahim El Bitar, Fatima-Zahra Belouadha, Ounsa Roudiès

Siweb team (Information System and Web), SIR laboratory
Ecole Mohammadia d'Ingénieurs, Mohammed V - Agdal University, Morocco.
{elbitar, belouadha, roudies}@emi.ac.ma

*Abstract*— **Most works on Web services has focused on discovery, composition and selection processes of these kinds of services. Other few works were interested in how to represent Web services search queries. However, these queries cannot be processed by ensuring a high level of performance without being adequately represented first. To this end, different query languages were designed. Even so, in the absence of a standard, these languages are quite various. Their diversity makes it difficult choosing the most suitable language. In fact, this language should be able to cover all types of preferences or requirements of clients such as their functional, non-functional, temporal or even specific constraints as is the case of geographical or spatial constraints and meet their needs and preferences helping to provide them the best answer. It must also be mutually simple and imposes no restrictions or at least not too many constraints in terms of prior knowledge to use and also provide a formal or semi-formal queries presentation to support their automatic post-processing. To help assisting developers in their choices and also identifying the criteria that should satisfy a standard language, this article presents a taxonomy of different classes of query languages designed to be able to express Web services queries. An illustrative example is also given to illustrate clearly the different query representations that can be offered by these languages classes. A comparative study is eventually established to allow to reveal the advantages and limitations of various existing languages in this context. It is a synthesis of this category of languages discussing their performance level and their capability to respond to various needs related to the Web services research and discovery case. The criterions identified at this stage may, in our opinion, constitute then the main pre-requisite that a language should satisfy to be called perfect or to be a future research standard.**

## I. INTRODUCTION

The paradigm of Web services, using standards based on XML (eXtensible Markup Language), is attracting the interest of many researchers, providers and users of the Web. As a result, their ever-increasing number continues evolving rapidly in order to satisfy different customer needs. Automating the Web services discovery of existing atomic Web services as well as the Web services composition by combining different features is then required. However, any automatic processing of clients needs require, first, to give them the means to write their requests before triggering the necessary treatments. Most works on Web services has focused on discovery, composition and selection processes of these kinds of services. Other few works were interested in how to represent Web services search queries. However, these queries cannot be processed by ensuring a high level of performance without being adequately represented first. Firstly, these queries have to express clearly the specific needs of clients, and secondly, be well represented in order to encourage and enable their automatic processing [1]. To this end, various search query languages for Web services were designed. These languages are the mean of communication that allows the client to use systems or platforms for Web services discovery and composition from the services repository, database or cache. They are quite diverse and differ both in techniques they adopt, as the goals they seek.

In this paper, we attempt to create a taxonomy of different languages for writing Web services search queries. The idea is to classify these languages on their approach to better understand and identify the advantages and disadvantages of each class. In this context, we illustrate the adopted query representation approaches of each class of the established taxonomy by a concrete example. The example we consider is a Web service for online airline ticket booking. This concrete study identifies a set of criteria to consider in comparing the different classes of languages we are concerned and provides a synthesis of their performance vis-a-vis the client's needs and Web service discovery and composition processes.

This paper is structured as follow. Section 2 reveals our research problematic and objectives. Section 3 presents a taxonomy of various related works, and subsequently, different query representational languages. Section 4 describes the different classes of concerned languages throughout an illustrative example. Section 5 provides a comparison of these languages based on a set of main criteria. Section 6 concludes this work and released a synthesis of the studied languages .

## II. PROBLEMATIC ET MOTIVATIONS

As we noted earlier, various languages for writing search Web services queries exist in the literature. They differ by both the techniques they adopt, as the goals they seek.

However, none of them has acquired the maturity to be the standard language.

Some languages do not cover all client needs, while others lack precision in the queries representation and eventually lead to performance problems on the achieved results or even they do not support automatic processing. We found that they are quite diverse and differ primarily by the adopted approach and its degree of response to a specific goal such as computer languages independence and usability, the query predisposition to continue its automatic processing, or the inclusion of its semantic aspect, or even the satisfaction of the client needs in terms of requirements and preferences. These requirements may in fact go beyond the functional frame and also target non-functional constraints such as quality of service (QoS) or the security level of the intended Web service.

Also, it is then important to think at first about how best to represent the client's query to better build up a system or a Web services discovery and composition platform. It would be necessary to answer some questions as: Which language would be the most suitable to take on, to meet the original goals of the system? Does the query representation which is adopted by the selected language, respond to other unforeseen challenges that may be a potential scalability of the system? Would this representation be directly interpreted by the machine or does it require a post-processing before calling the discovery or Web services composition process? Does it cover explicitly all the elements or needed knowledge to the Web services searching process or is it that they are implicit and must be deducted? Which impact would have the chosen representation on the obtained research results performance?

Our main motivation is to assist researchers in the Web services field, particularly, Web services discovery and composition platform developers, helping them to answer all these questions. The aim of this work is to study the different classes of existing Web services querying languages to reveal their advantages and disadvantages, and subsequently help developers in their choice. We believe that this study consents to identify a set of criteria that seems essential to be considered while future developing a standard language for representing the Web services search queries. There is no doubt that such language should satisfy different aspects of constraints: to be easy to use by non-computer scientist clients and to exceed the adopted query representations complexity, support the automatic processing of the client queries, meet their functional as well as non-functional different needs, and finally consider the semantics of the targeted domain to provide more precise queries, and subsequently lead to more efficient outcomes.

### III. TAXONOMY OF WEB SERVICES QUERYING LANGUAGES

The Web services query representation problem has been treated by several scientific communities seeking, firstly, to ensure client satisfaction, and second, to facilitate his queries automation processing. A detailed study of the various works done in this direction allowed us to classify the major languages which have been selected to represent or write Web services queries. This classification leads into a taxonomy consisting of four categories: the textual approach, the logical approach, the visual approach and the semantic approach.

The query languages category which is adopting the textual approach chooses to write queries in natural [2, 3] or semi-natural language [4]. Although natural language allows the client to easily express his inquiry, it does not provide a concrete, rather formal representation of the client's query as it is unable to unambiguously reflect its goals and needs. Semi-natural languages, for their part, are semi-formal languages that remain, however, close to the natural languages and therefore quite easy to use. For example, the query language SQL-XML [4] is based on a query interactive standard allowing the user to exploit XML documents through SQL (Standard Query Language) queries. In fact, the user expresses his query using SQL and gets XML data results. He must, therefore, be acquainted with the SQL to be able to properly express his own request.

The query languages category adopting the logical approach intends to allow the client to express various kinds of constraints on Web services such as non-functional ones. Many services can indeed respond to the same functional need without necessarily guaranteeing the same level of QoS or security as examples. These category related languages use different theories of logic such as temporal logic, propositional logic and first order logic to formulate client queries as constraints.

In this context, we refer to Karakoc and Senkul work where they proposed a logic based approach for the Web services automatic composition, responding to queries built on constraints [5]. Expression these constraints contribute to meet clients needs while taking into consideration their requirements in terms of QoS [5, 6, 7].

Otherwise, the query languages class adopting the visual approach opts for a graphical representation of queries. It allows the user to express his query as a graph. One of the most interesting works in this spot is the one who proposed the XML-GL [8]. It describes a model of graphical representation of the XML documents and DTDs (Document Type Definition) contents and defines a visual language for expressing queries on structured XML documents. The results are then encapsulated in structures such as lists or groups. The graphical representation of queries is a formal approach that facilitates especially their automatic processing.

Other studies have also used graphs to represent queries in quite specific and particular domains, such as geographic information systems. In this context, Egenhofer proposed a system that can express geographical type queries as topological configurations [9]. These queries can thus express, with specific graphs, location, position or objects orientation constraints. They can also express constraints on the spatial relationships or the distance between objects. Different relationships can be used to express by example when an object crosses another, covers it or is just close to it. On one hand, the use of a visual query language such as graphs reduces very significantly the complexity of the request since constraints will be expressed in a multidimensional way. On the other hand, the correspondence between the visual language elements and the constraints to express is direct. Mapping both of these options makes the language very intuitive [10].

Xlive is a lightweight mediator XML/XQuery that provides queries representations as tree graphs, specifically as tree model hyperlinked graphs called TGV (Tree Graph View) [11]. The mediator, developed at the University of Versailles, uses an engine based on an extension of XML algebra derived from the relational algebra to process data streams of XML trees. Its advantage is that it allows the Web services queries processing through the XQuery language in order to better optimize their execution plans. The TGV graph is in fact a more intuitive querying structure. It was particularly designed to allow direct optimization of a set of sub queries before generating their physical execution plan. This execution plan should eventually combine the results of all given sub queries to meet a specific need. It is generated by means of simple algebraic algorithms. The TGV graph is best qualified to translate, simplify and optimize queries. However, in order to optimize it, it is necessary to perform complex transformations to reduce the processed data and then simplify the resulting TGV graph.

The last query languages class which is adopting a semantic approach is summed up in OWL-QL (OWL Query Language). This language, used in the management of a request-response dialogue, is somehow establishing a protocol for both, a querying agent and a response agent who ought to communicate, seeking answers to Web services queries. OWL-QL is based on OWL (Ontology Web Language), just to be able to represent the desired knowledge semantics. According to Fikes and al., this language is much more qualified to be a draft standard querying language on the Semantic Web and constitute a request-response dialogue protocol between the client and the server agents of this field [12]. OWL-QL was in fact designed to be able to select appropriate services from a wide range of services and applications that seem to respond to a given query [13]. Although it is based on OWL, it constitutes a prototype which remains easily adaptable to use other languages or semantic models. OWL-QL is a formal specification that defines precisely the semantic relationships between the query, the answer and the knowledge base (KB) used to generate the response. Unlike standard databases and Web query languages, OWL-QL is not only based on the idea of request-response dialogue between the server and client in which the agent server can use specific methods of automated reasoning to generate responses to client requests, but also on dialogues with KBs that can be multiple, based on the Semantic Web concepts to access the required knowledge required to answer a query. Taking into account the semantic aspect is, indeed, a great contribution for efficient processing of queries. It is a major advantage that distinguishes OWL-QL from other approaches presented before.

IV.     ILLUSTRATIVE EXAMPLE

To illustrate more clearly the principles of the four Web services querying languages classes representations that we mentioned before, we consider an example of searching a Web service for online booking airline tickets. We assume that the client wants to find all existing Web services allowing him to book a plane ticket in business class for a flight from Casablanca to Beirut between August 1, 2011 and August 4, 2011 and costing less than 800 \$. We also assume that the customer prefers, if possible, that the departure time starts from 4 o'clock pm. This request cannot be reduced to a simple booking tickets Web services search. It is indeed quite complex because it incorporates several requirements and constraints related to client preferences. In this section, we will represent this query according to the different classes of approaches discussed above.

According to the textual approach, it is obvious that the same query could be expressed differently by different clients using a natural language. The text written by clients will certainly bring the same semantics but according to syntax, lexicons and different languages. Contrariwise, when it comes to using a semi-natural language such as XML-SQL, the query will have the form shown in Figure 1.

---

**Answer query:**

```
select  X.trip, X.price, Y.number Z.company, A.timeDep into
      Answer
from "trip.flight.price" X, "trip.flight.number" Y, "trip.flight.
      company" Z, "trip.flight.timeDep" A
where X.vol=Y.vol and X.vol=Z.vol and X.vol=A.vol
      and dbo.val(X.price)<800
group by flight.company
order by flight.price
```

**XML-SQL rule:**

```
<trip>
    <flight price="dbo.val(price)"> (flight)
    Company
    cDepart [dbo.val(cDepart)='Casablanca']
    cArrival [dbo.val(cArrival) = 'Beirut']
    dateDep [dbo.val(dateDep) <= '04/08/2011' &
    dbo.val(dateDep) >= '01/08/2011' ]
    class [dbo.val(class)= 'business']
    </flight>
</trip> :- Answer (flight, price, number, company, timeDep)
```

---

Figure 1.   Example of query using XML-SQL.

The user must express its SQL query using input variables which must then specify the required values in the form of rules and also output variables that allow it to retrieve the results. The provided query example, assumes that all travel information (flights) that are the outputs of the services discovered are described in XML documents whose structure is given in Figure 2. Indeed, if some Web services adopt outputs whose designation and structure are different, they will be ignored when searching for appropriate services even if they are actually able to answer the query in question. XML-SQL does not consider in fact the semantic aspect of the evolved knowledge.

```
<trip>
    <flight price= "740 $">
        <number>A15</number>
        <company>Qatar Airways</company>
        <cDepart>Casablanca</cDepart>
        <cArrival>Beirut</ cArrival >
        <dateDep>01/08/2011</dateDep>
        <dateArr>02/08/2011</dateArr>
        <timeDep>14.30</timeDep>
        <timeArr>****</timeArr>
        <class>business</class>
    </flight>
    < flight price= "...">
        ...
    </flight>
        ...
</trip>
```

Figure 2.   Example of structure of the XML documents informing on flights.

In the logical approaches case, we present in Figure 3 the same query using temporal logic. This consists on setting the variables and specifying their desired values and then expressing the set of conditions to check on these values.

Variables cDepart, cArrival, timeDep, dateDep, price, class
Init $\overset{\triangle}{=}$ cDepart = 'Casablanca' $\wedge$ cArrival = 'Beirut' $\wedge$ timeDep $\in$ (0..23)

**Condition to verifiy:**

$\exists$ (Init $\wedge$ (01/08/2011<=dateDep<=04/08/2011) $\wedge$ (price<800) $\wedge$(class= 'business') $\wedge \neg[]$ (timeDep>16.00))

Figure 3.   Example of query using temporal logic.

In the visual approaches case, we illustrate their graphs use to represent queries through the TGV graph. Figure 4 shows the use of the TGV tree graph to represent the query example we are considering in this article. The query is actually mapped into a graph whose three nodes are trees. Each tree of the source nodes (circled) is a sub query expressed by the variables it involves. The correspondence between the variables of each node (sub query) is indicated by hypertext links. It is primordial to express this correspondence based on accomplishing a join of two sub queries in response to the initial request. A special node (box) is the response node which is also a tree of variables constituting the answer. These variables are also linked to their corresponding nodes in the source (sub queries) where they get their values which are the results of executing the query.

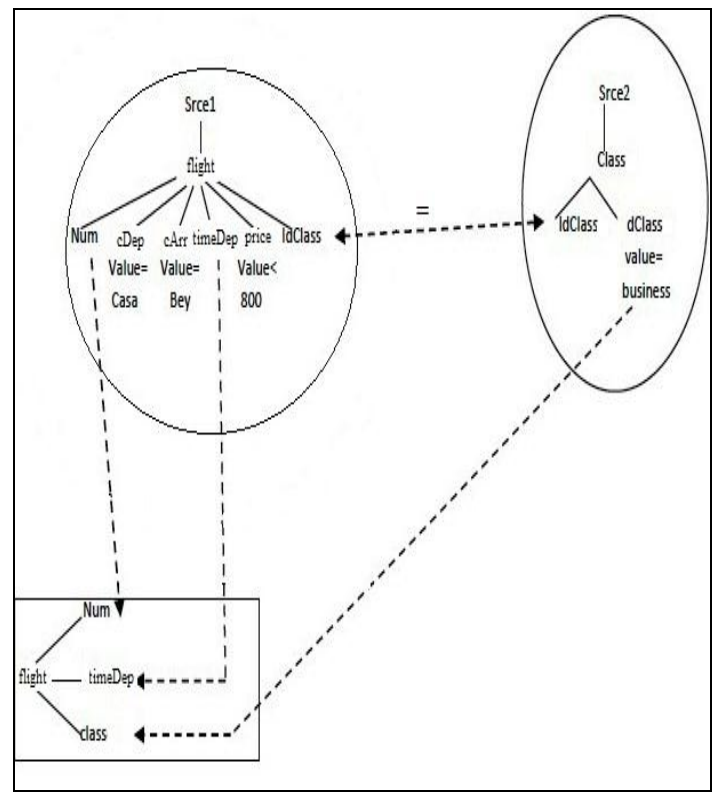

Figure 4.   Example of query using TGV.

Finally, in the semantic approach case, an OWL-QL request-response dialogue is initiated by a request sent by the client to the OWL-QL server. OWL-QL query is in fact an object necessarily containing a query pattern that specifies a collection of sentences respecting the OWL [12, 14]. This pattern, among others, includes URI references that are considered as URL variables used, in most cases, to access the concerned KBs (knowledge bases) or to communicate with their servers. Furthermore, our example query is expressed in OWL-QL as shown in Figure 5. This involves defining the given query in terms of involved variables and the desired results according to specific patterns.

---

**Query**: ("plane ticket in business class for a flight from Casablanca to Beirut between August 1, 2011 and August 4, 2011 and costing less than 800 $. We also assume that the customer prefers, if possible, that the departure time starts from 4 o'clock pm.")

**Query Pattern**: {(offer ?o ?f) (01/08/2011<dateD ?f <04/08/2011) (cityD ?f casablnanca) (cityA ?f beyrouth) (class ?c business) (price ?f <800) }
**Must-Bind Variables List**: (?o)
**May-Bind Variables List:** ()
**Don't-Bind Variables List:** ()
**Answer Pattern**: {(offre ?o " plane ticket in business class for a flight from Casablanca to Beirut between August 1, 2011 and August 4, 2011 and costing less than 800 $. We also assume that the customer prefers, if possible, that the departure time starts from 4 o'clock pm.")}

**Answer KB Pattern**: …

**Answer**: ("Qatar Airways offers plane ticket in business class for a flight from Casablanca to Beirut between August 1, 2011 and August 4, 2011 and costing less than 800 $. We also assume that the customer prefers, if possible, that the departure time starts from 4 o'clock pm.?")

**Answer Pattern Instance**: {(offer Qatar Airways " plane ticket in business class for a flight from Casablanca to Beirut between August 1, 2011 and August 4, 2011 and costing less than 800 $. We also assume that the customer prefers, if possible, that the departure time starts from 4 o'clock pm.")}

**Query**: …
**Server**: …

---

Figure 5. Example of query using OWL-QL.

## V. COMPARISON OF THE IDENTIFIED QUERYING LANGUAGES CLASSES

As discussed in previous sections, we identified four classes of Web services querying languages. To compare these classes, we first established the set of criteria which our comparison relies on. These criteria were derived from each class characteristics to respond to different needs that we have to satisfy with a Web services querying language. We are talking about the complexity, the degree of formality, the automaticity, the expressiveness, the constraints weighting and the semantic coverage.

The criterion of complexity reflects the ease of use. It refers to the constraint of whether or not to have perquisite knowledge to use the language. The degree of formality is a criterion that indicates that the being considered class applies a formal language used to represent the query in a non-ambiguous way and promotes its processing by the machine. The automaticity is a criterion that indicates the predisposition of the query expressed by the considered language to continue automatic treatment or the need for a post treatment consisting in converting it into a suitable form enabling such automatic processing. The expressivity is a criterion that indicates the ability to cover and to express different types of clients' needs in terms of functional requirements and preferences as well as non-functional, temporal or specific ones, as is the case of geographical or spatial constraints. In addition, the constraints weighting is a criterion indicating the ability of the language to allow the client to represent constraints with different weights, consequently distinguishing its demands from its preferences. For example, unlike all other languages, temporal logic distinguishes the departure time as preference and not as a constraint to be necessarily verified. Finally, the criterion of semantic coverage reflects the possibility of representing the semantics of the client's needs in the query.

To put side by side all the languages studied above the different established criteria, we have evaluated them on a scale of 0 to 2. We evaluate each criterion by 0 when the language does not suit at all the given criterion and by 2 when it is completely satisfied. Values between 0 and 2 were used to assess criteria that are partially satisfied by the language concerned. The values we have assigned to each language have been identified based on the descriptions of these languages in the existing works [2, 4, 8, 12, 15, 16, 17, 19]. We also treated fairly all the considered criteria while assigning the same weight. We believe, in fact, they are all important for the choice of a suitable Web services search querying language. Figures 6 and 7 summarize in a columns chart and a radar chart, the results of our studied languages criteria-comparison.

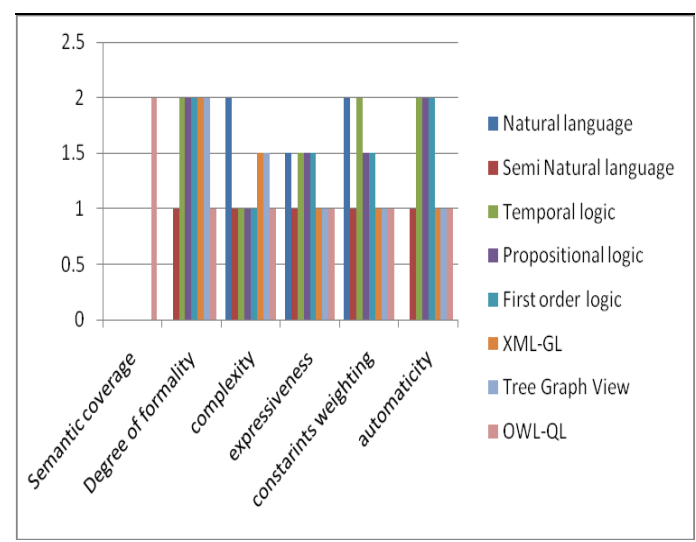

Figure 6. Columns chart comparing Web services querying languages.

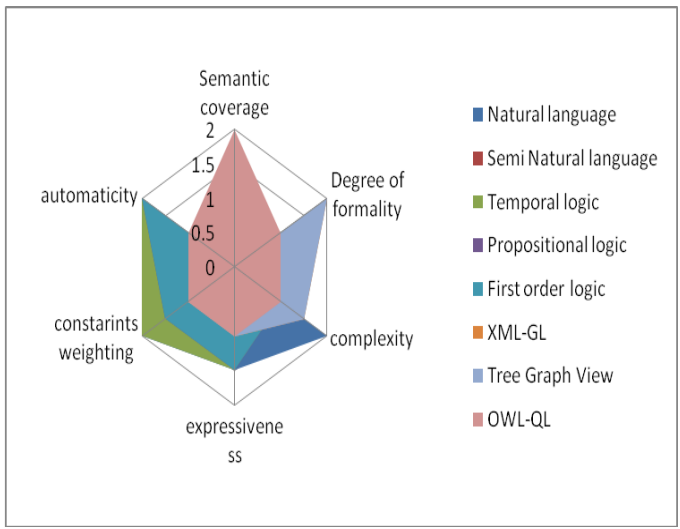

Figure 7.  Radar chart comparing Web services querying languages.

From the drawn graphs, we find that the languages of the textual class are generally the most appropriate in cases where queries must be raised by a simple user who is not expected to have a specific level of knowledge to write its query. As for languages based on logic, in particular temporal logic, they are formal specifications that support automatic processing. However, their strength lies in their ability to distinguish the client's requirements from his preferences. These two types of constraints are not treated the same. Moreover, the visual class languages, based on the use of graphs, have particularly the advantage of representing an intuitive application but also for representing execution plans when it comes to composing services in response to an initial request. The graph TGV, for example, can dissect a query into sub queries and then simply combine their results to meet a complex need. Finally, the semantic class represented by the OWL-QL, marks essentially the advantage of involving the semantic aspect of knowledge brought in the query. This aspect stills important to automate the discovery or Web services composition process.

## VI. Conclusion et synthesis

In this paper, we tried, by reviewing of the knowledge domain concerning writing Web services search queries, to establish a taxonomy of different languages used for this purpose. The four classes of this taxonomy were identified according to the approaches and the techniques they adopt for the representation of such queries. We could then point out the benefits as well as the limits of the languages associated to these classes. Concluding the established synthesis at the end of this work, we find that all of these languages  solve, in fact, only partially the issue of representation of queries related to Web services. No class has a perfect language. Indeed, no language, no matter its class, covers all necessary aspects to express queries that are simple to make, fairly formal to be processed automatically and taking into account all types of constraints required by the client. We believe that a standard language respecting the different criteria noted in Section 5 could be great helpful when it comes to discovery, composition, and also automatic selection of the best suited Web services, to better meet the specific needs of a client. Such language could possibly be inspired from existing approaches, but should have a number of qualities such as being expressive to consider and treat different types of constraints such as time constraints, functional or non-functional constraints, having a simple syntax or rather overcome its syntax complexity by providing an easy graphical notation to be used by a simple user, being independent of the used platforms, being formal or at least semi-formal to describe queries that can be processed automatically, being scalable to adapt easily to particular application domains such as geographic information systems, and preferably also covering the semantic aspect of queries to avoid ending up in inaccurate search results due to the semantic ambiguity problems of the client's real requirements and constraints.